\documentclass[a4paper,10pt]{article}

\title{Stochastic cosmology, theories of perturbations and Lifshitz gravity}
\author{I M Khalatnikov$^{1}$ and A Yu Kamenshchik$^{1,2}$}
\date{}
\begin{document}

\maketitle
\hspace{-5mm}$^1$L.D. Landau Institute for Theoretical Physics of Russian Academy of Sciences,
Kosygin str. 2, 119334 Moscow, Russia\\
$^2$Dipartimento di Fisica e Astronomia, Universit\`a di Bologna  and Istituto Nazionale di Fisica Nucleare, via Irnerio 46, 40126 Bologna, Italy
\begin{abstract}
We review some works of E M Lifshitz connected with gravity and cosmology and also
some later works, connected with his ideas. The main topics of this review are the stochastic cosmology of an anisotropic universe and of an isotropic universe with the scalar field, the quasi-isotropic (gradient) expansion in cosmology and Horava-Lifshitz gravity and cosmology.
\end{abstract}

\section{Introduction}

The name of Evgenii Mikhailovich Lifshitz is known to the physicists of all the world 
and even to more wide public in connection  with the famous course of theoretical physics 
written by him in collaboration with his friend and teacher Lev Davidovich Landau. 
In this review, dedicated to the 100th anniversary of the birth of E.M. Lifshitz we would like to dwell on some directions of theoretical physics, in fields of gravity and cosmology, developed by him and other members of the Landau school. Besides, we would like to speak about 
some unexpected trends in the modern physics, connected with his name.  

First of all, we shall discuss the stochastic or chaotic cosmology. As is well known, 
the  study of chaos has become very popular in modern physics and chaotic phenomena 
were found in different fields and, sometimes, in rather  simple systems (see e.g. \cite{chaos}). It is amusing that one of the first examples of  stochastic or chaotic 
behavior was discovered in the realm of cosmology. Initially the study of chaos in cosmology 
was connected with the investigation of the so called initial singularity problem. Later it has become clear that the stochastic properties of cosmological models  are not necessarily related to the singularity and can arise already in such simple systems as Friedmann universes.

To begin with, let us remember that R Penrose and S Hawking \cite{Pen-Hawk} proved
the impossibility of indefinite continuation of geodesics under certain conditions.
This was interpreted as pointing to the existence of a singularity in the general
solution of the Einstein equations. These theorems, however, did not allow for finding
the particular analytical structure of the singularity.
The analytical behaviour of the general solutions of the Einstein equations in the
neighborhood of singularity was investigated in the papers by E M Lifshitz and I M Khalatnikov
\cite{LK0,LK,LK1} and V A Belinsky, E M Lifshitz and I M Khalatnikov \cite{BKL,BKL1,BKL2}.
These papers revealed the enigmatic  phenomenon of an oscillatory approach to the singularity
which has become known also as the {\it Mixmaster Universe} \cite{Misner}.
The model of the closed homogeneous but anysotropic universe with three degrees of freedom
(Bianchi IX cosmological model) was used to demonstrate that the universe approaches the singularity
in such a way that its contraction along two axes is accompanied by an expansion with respect to
the third axis, and the axes change their roles according to a rather complicated law which reveals
chaotic behavior \cite{BKL1,BKL2,LLK,five}.

The study of the dynamics of the universe in the vicinity of the cosmological singularity
has become a rapidly developing field of modern theoretical and mathematical physics.
First of all we would like to mention the generalization of the study of the oscillatory approach
to the cosmological singularity in multidimensional cosmological models. It was noticed \cite{multi}
that the  approach to the cosmological singularity in the multidimensional (Kaluza-Klein type)
cosmological models has a chaotic character in the spacetimes whose dimensionality is not higher then
ten, while in the spacetimes of higher dimensionalities a universe after undergoing a finite number
of oscillations enters into monotonic Kasner-type contracting regime.

The development of cosmological studies based on  superstring models has revealed some new aspects
of the dynamics in the vicinity of the singularity \cite{DHN}.
First, it was shown that in these models exist mechanisms of changing of Kasner epochs provoked not by
the gravitational interactions but by the influence of other fields present in these theories.
Second, it was proved that the cosmological models based on six main superstring models plus $D=11$ supergravity
 model exhibit the chaotic oscillatory approach towards the singularity.
Third, the connection between cosmological models, manifesting the oscillatory approach towards singularity and
a special subclass of infinite-dimensional Lie algebras \cite{Kac} - the so called hyperbolic Kac-Moody algebras was
discovered.

Meanwhile, the cosmology at the edge of the eighties has experienced a new revolution, connected with the birth of the so called inflationary cosmology \cite{inflation}. 
The theory of a hot universe, which has begun its evolution from the singularity called Big Bang \cite{hot} was in a good correspondence with the observational data, especially 
after the discovery of the cosmic microwave background radiation in sixties by Penzias and Wilson, but it suffered from some fundamental problems, like those of flatness, of homogeneity, of horizon and some others. The theory of cosmological inflation, developed to resolve these problems, was based on the assumption that at the beginning of the cosmological evolution the universe had undergone a period of a quasi-exponential expansion. Such an expansion 
can be provided by the presence of an effective cosmological constant. However, to go out from the inflationary stage of the cosmological evolution, which should be rather short, it is necessary to provide the slow evolution of this effective constant, which is not constant. 
Thus, the models of Friedmann homogeneous and isotropic universe, filled with a scalar field, usually called inflaton become popular. 

The dynamics of such models was studied in detail in papers \cite{BGKZ, BK-scal} by using the methods of qualitative theory of differential equations.   
It was clear that a closed contracting Friedmann universe, filled with a massive scalar field can, sometimes, escape the falling to the singularity and  undergoes a bounce. In paper by D Page \cite{Page} the suggestion that the infinitely bouncing   trajectories in the closed 
Friedmann universe, filled with a scalar field constitute a fractal set was analyzed. In such a 
way the stochastic behavior was investigated in the framework of very simple cosmological models. The hypothesis of Page was further studied in papers \cite{KKT, KKT1} and further arguments in favor of the stochasticity of dynamics of Friedmann models were found. 

There was also another question, discussed in the context of studying of the stochastic behavior in cosmology -- the search for some invariant characteristics of the chaoticity. 
As is well known the general relativity is a reparametrization invariant theory and it important 
to have a coordinate system - independent quantities at studying of gravitational phenomena.
One of the invariant tools, useful for the 
studying of chaoticity in cosmology is the so called topological entropy. In paper \cite{CSh}
this tool was applied to the study of the simplest model of closed Friedmann universe filled with the scalar field, whose potential includes only the massive term. In paper \cite{CL} the topological entropy and some other characteristic were applied to the study of the Mixmaster Universe while in paper \cite{KKST} the topological entropy was calculated for a family of more compacted Friedmann models. 

If one would like to consider the cosmology not as a purely mathematical application of the general relativity and has an intention to compare its predictions with observational data, 
one cannot limit itself by studying the Friedmann or Bianchi universes. Thus, to treat the 
cosmological solutions of Einstein equations, we need to develop some perturbative methods. It is remarkable that the first paper devoted to the development of the theory of linear cosmological perturbations on the Friedmann backgrounds was written by E M Lifshitz as early as in  1946 \cite{Lif1946}. Then this theory was further developed in \cite{LK0}.
Later, a huge amount of work was done in this field. In particular, one should note the gauge-invariant theory of cosmological perturbations \cite{Bardeen, MBF, Sassaki}. 

While the theory of linear cosmological perturbations is basically applied to the Friedmann backgrounds, another asymptotic tool, developed by E M Lifshitz and I M Khalatnikov 
in paper \cite{quasi} - the so called quasi-isotropic expansion for the solutions of Einstein equations near singularity can take as a zero approximation more large class of backgrounds. This expansion was also largely studied, some times under the name 
of ``gradient expansion''  \cite{KKS, KKMS, KKS1}. In some sense this expansion can be
generalized also for the whole universe, without limitation to the vicinity of the singularity.  

Recently a new phenomenon of the cosmic acceleration was discovered \cite{cosmic}. 
This discovery  has stimulated the search for the so called dark energy, which can explain this phenomenon \cite{dark}.   An alternative or complementary way of the explanation of the cosmic acceleration is the construction of some modified theory of gravity. 
Besides, the quantum theory of gravity is non-renormalizable and this represents another stimulus for the modification of gravity.
Amongst different theories of modified gravity there is rather unexpected apparition - the Horava- Lifshitz theory of gravity \cite{Hor-Lif}, which is inspired by a couple of old papers by Lifshitz devoted to  phase transitions theory \cite{Lif-phase}. The main idea consists in the hypothesis that the graviton propagator can be Lorentz - non-invariant at big values of the momentum
and its spatial part behaves differently with respect to the temporal part. In such a way we can obtain the renormalizability of the theory without encountering such unpleasant objects as tachyons or ghosts. 
In spiteof the fact that the breaking the Lorentz invariance looks for many researchers as a too radical step, it is amusing how the old papers by Lifshitz, devoted to an object which stayed far away from the gravitational thematics, have provoked the creation of a new fashion in such a modern field as quantum gravity.

The structure of the paper is the following:
in Sec. 2 we shall recall the main features of the oscillatory approach to the singularity in relativistic cosmology and dwell on its stochastic nature;
Sec. 3 will be devoted 
to the study of the stochasticity in Friedmann cosmology;
in section 4 we present some new developments in the study of quasi-isotropic expansions;
in section 5 we shall give a brief review of the Horawa-Lifshitz gravity and  the last section is devoted to the conclusions.

\section{Oscillatory approach to the singularity and stochastic 
cosmology}
One of the first exact solutions found in the framework of general
relativity was the Kasner solution \cite{Kasner} for the Bianchi-I
cosmological model representing gravitational field in an empty space
with Euclidean metric depending on time according to the formula
\begin{equation}
ds^{2} = dt^{2} - t^{2p_{1}}dx^{2} - t^{2p_{2}}dy^{2}
-t^{2p_{3}}dz^{2},
\label{Kasner}
\end{equation}
where the exponents $p_{1}, p_{2}$ and $p_{3}$ satisfy the relations
\begin{equation}
p_{1}+ p_{2}+ p_{3} = p_{1}^{2}+ p_{2}^{2}+ p_{3}^{2} = 1.
\label{exponents}
\end{equation}
Choosing the ordering of exponents as
\begin{equation}
p_{1} < p_{2} < p_{3}
\label{ordering}
\end{equation}
one can parameterize them as \cite{LK0}
\begin{equation}
p_{1} = \frac{-u}{1+u+u^{2}},\;
p_{2} = \frac{1+u}{1+u+u^{2}},\;
p_{3} = \frac{u(1+u)}{1+u+u^{2}}.
\label{u-define}
\end{equation}
As the parameter $u$ varies in the range $u \geq 1$,
$p_{1}, p_{2}$ and $p_{3}$ assume all their permissible values
\begin{equation}
-\frac{1}{3} \leq p_{1} \leq 0,\; 0 \leq p_{2} \leq \frac{2}{3},\;
\frac{2}{3} \leq p_{3} \leq 1.
\label{range0}
\end{equation}
The values $u < 1$ lead to the same range of values of
$p_{1},p_{2},p_{3}$ since
\begin{equation}
p_{1}\left(\frac{1}{u}\right) = p_{1}(u),\;
p_{2}\left(\frac{1}{u}\right) = p_{3}(u),\;
p_{3}\left(\frac{1}{u}\right) = p_{2}(u).
\label{u-define1}
\end{equation}
The  parameter $u$ introduced in early sixties has
appeared very useful and its properties attract attention of researchers
in different contests. For example, in the recent paper \cite{LK-Pet} a connection was
established between the Lifshitz-Khalatnikov parameter $u$ and the invariants, arising
in the context of the Petrov's classification of the Einstein spaces \cite{Petrov}.

In the case of Bianchi-VIII or Bianchi-IX cosmological models
the Kasner regime (\ref{Kasner}), (\ref{exponents}) ceased to be an
exact solution of Einstein equations, however one can design the
generalized Kasner solutions \cite{LK}-\cite{BKL2}.
It is possible to construct some kind of perturbation theory where
exact Kasner solution (\ref{Kasner}), (\ref{exponents}) plays role of
zero-order approximation while the role of perturbations play those terms
in Einstein equations which depend on spatial curvature tensors
(apparently, such terms are absent in Bianchi-I cosmology). This
theory of perturbations is effective in the vicinity of singularity
or, in other terms, at $t \rightarrow 0$. The remarkable feature of
these perturbations consists in the fact that they imply the
transition from the Kasner regime with one set of parameters
to the Kasner regime with another one.

The metric of the generalized Kasner solution in a synchronous
reference system can be written in the form
\begin{equation}
ds^{2} = dt^{2} - (a^{2}l_{\alpha}l_{\beta}
+b^{2}m_{\alpha}m_{\beta}+c^{2}n_{\alpha}n_{\beta})dx^{\alpha}
dx^{\beta},
\label{metric}
\end{equation}
where
\begin{equation}
a = t^{p_{l}},\;b = t^{p_{m}},\;c = t^{p_{n}}.
\label{exponents1}
\end{equation}
The three-dimensional vectors $\vec{l}, \vec{m}, \vec{n}$ define the
directions along which the spatial distances vary with time according
to the power laws (\ref{exponents1}).
Let $p_{l} = p_{1}, p_{m} = p_{2}, p_{n} = p_{3}$ so that
\begin{equation}
a \sim t^{p_{1}},\;b \sim t^{p_{2}},\;c \sim t^{p_{3}},
\label{exponents2}
\end{equation}
i.e. the Universe is  contracting in directions given by vectors
$\vec{m}$ and $\vec{n}$ and is expanding along $\vec{l}$.
It was shown  that the perturbations caused by
spatial curvature terms make the variables $a, b$ and $c$ to undergo
transition to another Kasner regime characterized by the following
formulae:
\begin{equation}
a \sim t^{p_{l}'},\;b \sim t^{p_{2}'},\;c \sim t^{p_{3}'},
\label{Kasner1}
\end{equation}
where
\begin{equation}
p_{l}' = \frac{|p_{1}|}{1-2|p_{1}|},\;
p_{m}' = -\frac{2|p_{1}|-p_{2}}{1-2|p_{1}|},\;
p_{n}' = -\frac{p_{3}-2|p_{1}|}{1-2|p_{1}|}.
\label{Kasner2}
\end{equation}

Thus, the effect of the perturbation is to replace one ``Kasner
epoch'' by another so that the negative power of $t$ is shifted
from the $\vec{l}$ to the $\vec{m}$ direction. During the transition
the function $a(t)$ reaches a maximum and $b(t)$ a minimum. Hence,
the previously decreasing quantity $b$ now increases, the quantity
$a$ decreases and $c(t)$ remains a decreasing function. The
previously increasing perturbation caused the transition from regime
(\ref{exponents2}) to that (\ref{Kasner1}) is damped and eventually
vanishes. Then other perturbation begins grow which leads to a new
replacement of one Kasner epoch by another, etc.

We would like to emphasize  that namely the fact that perturbation
implies such a change of dynamics which extinguishes it, gives us an
opportunity to use perturbation theory so successfully. Let us note
that the effect of changing of the Kasner regime exists already in
the cosmological models more simple than those of Bianchi IX and
Bianchi VIII. As a matter of fact in the Bianchi II universe there
exists only one type of perturbations, connected with the spatial
curvature and this perturbation makes one change of Kasner regime
(one bounce). This fact was known to E M Lifshitz and I M
Khalatnikov at the beginning of sixties and they have discussed this
topic with L D Landau (just before the tragic accident) who has
appreciated it highly. The  results describing  the dynamics of the
Bianchi IX model were reported by I M Khalatnikov in his talk given
in January 1968 in Henri Poincare Seminar, in Paris. John A Wheeler
who was present there pointed out that the dynamics of the Bianchi
IX universe represents a non-trivial example of the chaotic
dynamical system. Later Kip Thorn has distributed the preprint with
the text of this talk.

Coming back to the rules governing  the bouncing of the negative
power of time from one direction to another one can show that they
could be conveniently expressed by means of the parameterization
(\ref{u-define}):
\begin{equation}
p_{l} = p_{1}(u),\; p_{m} = p_{2}(u),\; p_{n} = p_{3}(u)
\label{u-define2}
\end{equation}
and then
\begin{equation}
p_{l}' = p_{2}(u-1),\; p_{m}' = p_{1}(u-1), \;p_{n}' = p_{3}(u-1).
\label{u-define3}
\end{equation}
The greater of the two positive powers remains positive.

The successive changes (\ref{u-define3}), accompanied by a bouncing
of the negative power between the directions $\vec{l}$ and $\vec{m}$,
continue as long as the integral part of $u$ is not exhausted,
i.e. until $u$ becomes less that one. Then, according to  Eq. (\ref
{u-define1}) the value $u < 1$ transforms into $u > 1$, at this
moment either the exponent $p_{l}$ or $p_{m}$ is negative and $p_{n}$
becomes smaller one of the two positive numbers ($p_{n} = p_{2}$).
The next sequence of changes will bounce the negative power between
the directions $\vec{n}$ and $\vec{l}$ or $\vec{n}$ and $\vec{m}$.
Let us emphasize that the usefulness of the Lifshitz-Khalatnikov
parameter $u$ is connected with the circumstance that it allows
to encode rather complicated laws of transitions between different
Kasner regimes (\ref{Kasner2}) in such simple rules as
$u \rightarrow  u - 1$ and
 $u \rightarrow  \frac{1}{u}$.

Consequently, the evolution of our model towards a singular point
consists of successive periods (called eras) in which expansions and contractions
 of scale factors along
two axes oscillate while the scale factor along the third axis decreases monotonically,
the volume decreases according to a law which is near to $\sim t$. In
the transition  from one era to another, the axes along which the
distances decrease monotonically are interchanged. The order in which
the pairs of axes are interchanged and the order in which eras of
different lengths follow each other acquire a stochastic character.

To every ($s$th) era corresponds a decreasing sequence of values of
the parameter $u$.  This sequence has the form $u_{max}^{(s)},
u_{max}^{(s)}-1,\ldots,u_{min}^{(s)}$, where $u_{min}^{(s)} < 1$.
Let us introduce the following notation:
\begin{equation}
u_{min}^{(s)} = x^{(s)},\; u_{max}^{(s)} = k^{(s)} + x^{(s)}
\label{era}
\end{equation}
i.e. $k^{(s)} = [u_{max}^{(s)}]$ (the square brackets denote
the greatest integer $\leq u_{max}^{(s)}$). The number $k^{(s)}$
defines the era length. For the next era we obtain
\begin{equation}
u_{max}^{(s+1)} = \frac{1}{x^{(s)}},\;
k^{(s+1)} = \left[\frac{1}{x^{(s)}}\right].
\label{era1}
\end{equation}

The ordering with respect to the length of $k^{(s)}$ of the
successive eras (measured by the number of Kasner epochs contained in
them) acquires asymptotically a stochastic character . The random
nature of this process arises because of the rules
(\ref{era})--(\ref{era1}) which define the transitions from one era
to another in the infinite sequence of values of $u$. If all these
infinite sequences begin from some initial value $u_{max}^{(0)}
= k^{(0)} + x^{(0)}$, then the lengths of series $k^{(0)},
k^{(1)},\ldots$ are numbers included into an expansion of a
continuous fraction:
\begin{equation}
k^{(0)} + x^{(0)} = k^{(0)} + \frac{1}{k^{(1)} +
\frac{1}{k^{(2)}+\cdots}} .
\label{fraction}
\end{equation}

We can describe statistically this sequence
of eras if we consider instead of a given initial value
$u_{max}^{(0)} = k^{(0)} + x^{(0)}$ a distribution of $x^{0)}$ over
the interval $[0,1]$ governed by some probability law. Then we also
obtain some distributions of the values of $x^{(s)}$ which terminate
every $s$th series of numbers. It can be shown that with increasing
$s$, these distributions tend to a stationary (independent of $s$)
probability distribution $w(x)$ in which the initial value $x^{(s)}$
is completely ``forgotten'':
\begin{equation}
w(x) = \frac{1}{(1+x) \ln 2}.
\label{distrib}
\end{equation}
It follows from Eq. (\ref{distrib}) that the probability distribution
of the lengths of series $k$ is given by
\begin{equation}
W(k) = \frac{1}{\ln 2} \ln \frac{(k+1)^{2}}{k(k+2)}.
\label{distrib1}
\end{equation}

Moreover, one can calculate in an exact manner probability
distributions for other parameters describing successive
eras such as parameter $\delta$ giving relation between the
amplitudes of logarithms of functions $a, b, c$ and logarithmic time
\cite{five}. Thus, let us describe briefly also this further development of 
the description of the cosmological stochasticity, presented in the paper 
by I M Khalatnikov, E M Lifshitz, K M Khanin, L N Shchur and Ya G Sinai 
\cite{five}. 

First, it was noticed that from a formal point of view one encounters here  
a deterministic dynamical
model, governed by a system of three ordinary differential equations (the spatial-spatial components of the Einstein equations) plus an additional condition (the time-time component of the Einstein equations): thus 
 the phase space of this system is
actually not six but five dimensional. Thus, apart from the actual profound
cosmological significance of this system, we have encountered here a specific mode
of spontaneous stochastization of a deterministic system.

Then, it was underlined that the transition from one Kasner era to another  can be described by the mapping transformation of the interval $[0,1]$ into itself by the formula 
\begin{equation}
 Tx = \left\{\frac{1}{x}\right\},\ \ x_{s+1} = \left\{\frac{1}{x_{s}}\right\},
 \label{Gauss0}
 \end{equation}
where curly brackets stay for the fractional part of a number. 
This transformation belongs to the so-called expanding transformations
of the interval $[0, 1]$, i.e., transformations $x\sim f(x)$ with $|f'(x)| > 1$. Such
transformations possess the property of exponential instability: if we take
initially two close points, their mutual distance increases exponentially under
the iterations of the transformations. It is well known that the exponential
instability leads to the appearance of strong stochastic properties.

To study quantitatively the stochastic properties of the transitions between the Kasner eras
it is convenient to introduce some new notations. 
The logarithmic time is 
\begin{equation}
\Omega = -\ln t.
\label{time-log}
\end{equation}
The parameters $\alpha,\beta$ and $\gamma$ are the logarithms of the scale factors:
\begin{equation}
\alpha = \ln a,\ \beta = \ln b,\ \gamma = \ln c.
\label{scale-log}
\end{equation}
In what follows we shall discuss statistical properties of the sequence of
eras. The index $s$ numbers eras beginning from an arbitrarily chosen initial
one ($s = 0$). The symbol $\Omega_s$ denotes the initial instant of the $s$th era (defined
as the instant when the scale function, which was monotonically decreasing
during the preceding era begins to increase). The initial amplitudes of that
pair from among the functions $\alpha, \beta, \gamma$, which experiences oscillation in a given
era, we denote as $\delta_s\Omega_s$; the quantities $\delta_s$, (which assume values between 0 and
1) measure these amplitudes in units of the corresponding $\Omega_s$. The
recurrence formulas which determine the rules of transition from an era to
the next one are 
\begin{equation}
\frac{\Omega_{s+1}}{\Omega_s}=1+\delta_sk_s\left(k_s+x_s+\frac{1}{x_s}\right)\equiv \exp \xi_s,
\label{recur}
\end{equation}
\begin{equation}
\delta_{s+1}=1-\frac{\left(\frac{k_s}{x_s}+1\right)\delta_s}{1+\delta_sk_s\left(k_s+x_s+\frac{1}{x_s}\right)}.
\label{recur10}
\end{equation}
Iteration of this formula gives 
\begin{equation}
\frac{\Omega_s}{\Omega_0}=\exp\sum_{p=1}^s\xi_p.
\label{recur2}
\end{equation}

The quantities $\delta_s$ have a stable stationary statistical distribution $P(\delta)$ 
and a stable (small relative fluctuations) mean value.
This distribution $P(\delta)$ can be found exactly by an analytical method. 
Since we are interested in statistical properties in the stationary limit, it
is reasonable to introduce the so-called natural extension of the transformation
(\ref{Gauss0}) by continuing it without limit to negative indices. Otherwise
stated we pass over from a one-sided infinite sequence of the numbers
$(x_0,x_1,x_2,\ldots)$, connected by the equalities (\ref{Gauss0}), to a``doubly infinite''
sequence $X= (\ldots, x_i, x_0, x_l, x_2,\ldots)$ of the numbers which are connected by
the same equalities for all $-\infty < s < \infty$.

The sequence $X$ is equivalent to a sequence of integers $K=(\ldots,k_1,k_0,\ldots)$, constructed by the rule $k_s=[1/x_{s-1}]$. Inversely, every number of $X$ is determined 
by the integers of $K$ as an infinite continuous fraction 
\begin{equation}
x_s=\frac{1}{k_{s+1}+\frac{1}{k_{s+2}+\cdots}}\equiv x_{s+1}^+.
\label{recur3}
\end{equation}
The definition of $x_s^+$ can be written as 
\begin{equation}
x_s^+=[k_s,k_{s+1},\ldots].
\label{recur4}
\end{equation}
We also introduce the quantities which are defined by a continuous fraction
with a retrograde (in the direction of diminishing indices) sequence of
denominators:
\begin{equation}
x_{s}^-=[k_{s-1},k_{s-2},\ldots].
\label{recur5}
\end{equation}
We transform now the recurrence relation (\ref{recur10}) by introducing
 the notation $\eta_s=(1-\delta_s)/\delta_s$. Then (\ref{recur10}) can be rewritten as
\begin{equation}
\eta_{s+1}x_s=\frac{1}{\eta_s x_{s-1}+k_s}.
\label{recur6}
\end{equation}
By iteration we arrive at an infinite continuous fraction
\begin{equation}
\eta_{s+1}x_s=[k_s,k_{s-1},\ldots]=x_{s+1}^-.
\label{recur7}
\end{equation}
Hence $\eta_s=x_s^-/x_s^+$ and finally
\begin{equation}
\delta_s=\frac{x_s^+}{x_s^++x_s^-}.
\label{recur8}
\end{equation}

This expression for $\delta_s$ contains only two (instead of three) random quantities 
$x_s^+$ and $x_s^-$, each of which assumes values in the interval $[0,1]$. 

It follows from the definition (\ref{recur5}) that $1/x_{s+1}^-=x_s^-+k_s=x_s^-+[1/x_s^+]$.
Hence the shift of the entire sequence by one step to the right means a joint transformation 
of the quantities $x_s^+$ and $x_s^-$ according to 
\begin{equation}
x_{s+1}^+=\left\{\frac{1}{x_s^+}\right\},\ x_{s+1}^-=\frac{1}{\left[\frac{1}{x_s^+}\right]+x_s^-}.
\label{recur9}
\end{equation}
This is a one-to-one mapping in the unit square. Thus we have now a one-to-one
transformation of two quantities instead of a not one-to-one transformation
(\ref{Gauss}) of one quantity.
The quantities $x_s^+$ and $x_s^-$ have a joint stationary distribution
$P(x^+, x^-)$. Since (\ref{recur9}) is a one-to-one transformation, the condition for the
distribution to be stationary is expressed  by an  equation
\begin{equation}
P(x_s^+,x_s^-)=P(x_{s+1}^+,x_{s+1}^-)J,
\label{recur101}
\end{equation}
where $J$ is the Jacobian of the transformation. The normalized solution of this equation is 
\begin{equation}
P(x^+,x^-)=\frac{1}{(1+x^+x^-)^2\ln 2},
\label{recur-dis}
\end{equation}
its integration over $x^+$ or $x^-$ yields the function $w(x)$ (\ref{distrib}). 

Since by (\ref{recur8}) $\delta_s$ is expressed in terms of the random quantities $x_s^+$ and $x_s^-$, the knowledge of their joint distribution makes it possible to calculate the 
statistical distribution $P(\delta)$ by integrating $P(x^+,x^-)$ over one of thee variables at a constant value of $\delta$. 

Due to symmetry of the function (\ref{recur-dis}) with respect to the variables $x^+$ and $x^-$ we have $P(\delta)=P(1-\delta)$, i.e. the function $P(\delta)$ is symmetrical with respect to the point $\delta = 1/2$. We have 
\begin{equation}
P(\delta)d\delta=d\delta\int_0^1P\left(x^+,\frac{x^+\delta}{1-\delta}\right)\left(\frac{\partial x^-}{\partial \delta}\right)_{x^+}dx^+.
\label{recur-dis1}
\end{equation}
On evaluating this integral, we obtain finally
\begin{equation}
P(\delta)=\frac{1}{(|1-2\delta|+1)\ln 2}.
\label{delta-dis}
\end{equation}

The mean value $\langle\delta\rangle = 1/2$ already as a result of the symmetry of the
function $P(\delta)$. Thus, the mean value of the initial (in every era) amplitude of
oscillations of the functions $\alpha, \beta, \gamma$ increases as $\Omega/2$.

Thus, we have seen from the results of statistical analysis of
evolution in the neighborhood of singularity \cite{LLK,five} that
the stochasticity and probability distributions of parameters arise
already in classical general relativity.

At the end of this section a historical remark is in order. The
continuous fraction (\ref{fraction}) was shown in 1968 to I M
Lifshitz (L D Landau has already left) and he immediately noticed
that one can derive the formula for a stationary distribution of the
value of $x$ (\ref{distrib}). Later it becomes known that this
formula was derived in nineteenth century by Karl F  Gauss, who had
not published it, but had described it in a letter to one of
the colleagues.

\section{Stochastic Friedmann cosmology}

In the preceding section we have discussed the stochasticity in the Mixmaster universe.
However, already simple
isotropic closed Friedmann-Robertson-Walker models manifest
some elements of chaotic behavior which should
be taken into account for correct construction of quantum cosmological
theories  \cite{CSh}.

The studying of classical dynamics of a closed isotropic cosmological
model has a long story. First, it was noticed that in such a model
with a minimally coupled massive scalar field there is opportunity to
escape singularity at contraction \cite{Park-Ful, Star}. Then, the periodical
trajectories escaping singularity were studied \cite{Hawking}.
In paper \cite{Page} it was argued that  the set of infinitely bouncing
aperiodical trajectories has a fractal nature.  Later this result in
other terms was reproduced in our papers \cite{KKT,KKT1,KK,KK1}.

Here we would like
to describe briefly the approach presented in \cite{KKT}. The main idea
consisted in the fact that in the closed isotropic model with a
minimally coupled massive scalar field all the trajectories have the
point of maximal expansion. The localization of the points of maximal expansion
on the configuration plane $(a,\phi)$, where $a$ is a cosmological radius,
while $\phi$ is a scalar field could be found analytically. Then
the trajectories could be classified according to the localization of their
points of maximal expansion. The area of the points of maximal expansion
is located inside the so called Euclidean or ``classically forbidden''
region. Numerical investigations show that this area has a quasi-
periodical structure, zones corresponding to the falling to singularity
are intermingled with zones in which are placed points
of maximal expansion of trajectories having the so called ``bounce''
or point of minimal contraction. Then, studying the substructure
of these zones from the point of view of possibility to have two
bounces, one can see that this substructure repeats on the qualitative
level the structure of the whole region of the possible points of
maximal expansion.  Continuing this procedure {\it ad infinitum}
one can see that as the result one has the fractal set of infinitely
bouncing trajectories.

The same scheme gives us an opportunity to
see that there is also the set of periodical trajectories.
All these periodical trajectories contains bounces intermingled with
series of oscillations of the value of scalar field $\phi$.
It is important that there is no restrictions on the lengths of series
of oscillations in this case. In paper by Cornish and Shellard
\cite{CSh}  the topological entropy was calculated for this case
and it was shown that it is positive.

In paper \cite{KKST} the 
calculations of Ref. \cite{CSh}  were reproduced and it was shown how it was possible to
generalize them for more complicated cases.  
Here we shall briefly review the results of paper \cite{KKST}. 

First of all, let us write down the action for the simplest cosmological
model with the scalar field:
\begin{equation}
S = \int d^{4} x \sqrt{-g}\left\{\frac{m_{P}^{2}}{16\pi} (R -
2\Lambda) + \frac{1}{2} g^{\mu\nu}\partial_{\mu}\varphi
\partial_{\nu}\varphi -\frac{1}{2}m^{2}\varphi^{2}\right\},
\label{action}
\end{equation}
where $m_{P}$ is Planck mass, $\Lambda$ is cosmological constant.
Equations of motion for closed isotropic Universe are
\begin{equation}
\frac{m_{P}^{2}}{16 \pi}\left(\ddot{a} + \frac{\dot{a}^{2}}{2 a}
+ \frac{1}{2 a} \right)
+\frac{a \dot{\varphi}^{2}}{8}
-\frac{m^{2} \varphi^{2} a}{8}-\frac{m_{P}^{2}}{8 \pi}\Lambda a = 0,
\label{equation1}
\end{equation}
\begin{equation}
\ddot{\varphi} + \frac{3 \dot{\varphi} \dot{a}}{a}
+ m^{2} \varphi = 0.
\label{equation2}
\end{equation}
The first integral of motion of our system is
\begin{equation}
-\frac{3}{8 \pi} m_{P}^{2} (\dot{a}^{2} + 1)
+\frac{a^{2}}{2}\left(\dot{\varphi}^{2} + m^{2} \varphi^{2} +
\frac{m_{P}^{2}}{8 \pi} \Lambda \right)  =
0.
\label{integral}
\end{equation}
In the case when cosmological constant is equal to zero, the form of the
boundary of
Euclidean region is given by equation which can be easily obtained
from Eq. (\ref{integral}):
\begin{equation}
m^{2} a^{2} \phi^{2} = \frac{3}{4\pi}m_{P}^{2}.
\label{Euclid}
\end{equation}
Namely, in this model investigated in many papers \cite{Hawking,Page,
KKT,CSh} there are periodical trajectories with an arbitrary
number of oscillations of the scalar field.

Inclusion of positive cosmological constant, investigated in \cite{KKT1},
implies two opportunities: if $\Lambda$ is small in comparison with
the mass of the scalar field $m$, the qualitative behavior is the same as in the
model without cosmological constant; if cosmological constant is of order
$m^{2}$ the chaotic dynamics disappears in a jump-like manner \cite{KKT1}.

If we include into consideration hydrodynamical matter with the equation
of state
\begin{equation}
p = \gamma \epsilon,
\label{state}
\end{equation}
where $p$ is the pressure, $\epsilon$ is an energy density and $\gamma$ is
the constant ($\gamma = 0$ corresponds to dust matter, $\gamma = 1/3$
corresponds to radiation, while $\gamma = 1$ describes the massless
scalar field).
In this case, the form of the boundary of the Euclidean region is given
by the equation
\begin{equation}
m^{2} a^{2} \phi^{2} = \frac{3}{4\pi}m_{P}^{2} - \frac{D}{a^{q}},
\label{Euclid1}
\end{equation}
where $D$ is the constant characterizing the quantity of the given type
of matter in the Universe and $q = 3(\gamma+1) - 2$. 
Numerical calculations show that in this case again only restricted number
of oscillations is possible. Moreover, the structure of periodical
trajectories in this case is much more complicated. Indeed, the law is the
following: long series of oscillations can occur after a bounce followed by
a short series of oscillations while the behavior of a trajectory after a
short series of oscillations is less restricted.
The concrete laws, governing the structure of trajectories, depend on
the parameters of the model under consideration.
However, in this case
it is also  possible to calculate a topological entropy, which will be
demonstrated below.

Let us calculate the topological entropy for some of considered
cases. Topological entropy measures the growth in the number of periodical
orbits as their period increases.
Let us define \\
$N(k)$ - the number of periodic orbits of length $k$.\\
The definition of topological entropy looks as follows:
\begin{equation}
H_{T} = lim_{k \rightarrow \infty} \frac{1}{k} \ln N(k).
\label{definition}
\end{equation}
If $H_{T} > 0$, one can conclude that the dynamics is chaotic.

One can quantify the length of the orbit
by the number of symbols.
To begin with, let us reproduce in some detail the calculations from
Ref. \cite{CSh}. It was suggested to use the discrete coding of orbits,
containing  two symbols:\\
$A$ - the bounce of trajectory,\\
$B$ - crossing of line $\phi = 0$.\\
For the simplest model with the scalar field there is the only
prohibition rule: two letters $A$ cannot stay together, which means
that it is impossible to have two bounces one after another without
oscillations between them.

Let us denote by $Q(k)$ the number of ``words'' (trajectories) of length
 $k$ satisfying this rule, which begin with $A$ and end with $A$ and
by $P(k)$ the number of words which begins with $A$ and end with $B$.
Then one can easily write down  recurrent relations:
\begin{eqnarray}
&&Q(k+1) = P(k);\nonumber \\
&&P(k+1) = Q(k) + P(k).
\label{recurr}
\end{eqnarray}
It is easy to get from Eqs. (\ref{recurr})
the following relation for $P(k)$:
\begin{equation}
P(k+1) = P(k) + P(k-1).
\label{recurr1}
\end{equation}
One can easily calculate that
\begin{equation}
P(2) = 1, P(3) = 1
\label{initial}
\end{equation}
and that Eq. (\ref{recurr1}) defines the series of Fibonacci numbers.
Let us remember how to find the formula for the general term
of the Fibonacci series. One can  look for $P(k)$ as the linear
combination of terms $\lambda^{k}$, where $\lambda$ is the solution
of equation
$$
\lambda^{k+1} = \lambda^{k} + \lambda^{k-1}
$$
or, equivalently, because we are interested only in the non-zero roots
\begin{equation}
\lambda^{2} - \lambda - 1 = 0.
\label{equation}
\end{equation}
Looking for $P(k)$ in the form
\begin{equation}
P(k) = c_{1} \lambda_{1}^{k} + c_{2}\lambda_{2}^{k},
\label{form}
\end{equation}
where $\lambda_{1}$ and $\lambda_{2}$ are the roots of Eq.
(\ref{equation})
and satisfying the conditions (\ref{initial}), one can get
\begin{equation}
P(k) = \frac{1}{\sqrt{5}}\left[\left(
\frac{1+\sqrt{5}}{2}\right)^{k-1} + (-1)^{k-2}
\left(\frac{\sqrt{5}-1}{2}\right)^{k-1}\right].
\label{solution}
\end{equation}
Substituting (\ref{solution}) in the definition of the topological
entropy (\ref{definition}) one can find
\begin{equation}
H_{T} = \ln \left(\frac{1+\sqrt{5}}{2}\right) > 0,
\label{entropy}
\end{equation}
where
$\left(\frac{1+\sqrt{5}}{2}\right)$
is the famous golden mean. It is clear that only the largest root
of Eq. (\ref{equation}) is essential for the calculation of
the topological entropy.

Now one can go to more involved case of the cosmological model
with the scalar field and a negative cosmological constant. As it was described
above, in this model the periodical trajectories can have only a restricted
number of oscillations of the scalar field $\phi$. This rule can be
encoded in the prohibition to have more than $n$ letters $B$ staying
together, where the number $n$ depends on the parameters of the model. 

Now the recurrent relations look as follows:
\begin{eqnarray}
&&Q(k+1) = P(k)\nonumber \\
&&P(k+1) = Q(k) + P(k) - Q(k-n)\theta(k-n),
\label{recur}
\end{eqnarray}
where $\theta$ - function is defined in a usual manner.
We are interested in the limit $k \rightarrow \infty$ and
can substitute instead of $\theta(k-n)$ the number $1$. Now one can write
down the recurrent relation:
\begin{equation}
P(k+1) = P(k) + P(k-1) - P(k-n-1),
\label{recur1}
\end{equation}
which in turn implies
the following equation for the topological entropy:

\begin{equation}
\lambda^{n+2} - \lambda^{n+1} - \lambda^{n} + 1 = 0,
\label{equation1}
\end{equation}
where the topological entropy is equal to the logarithm of the biggest root
of Eq. (\ref{equation1}):
\begin{equation}
H_{T} = \ln \lambda.
\label{entropy1}
\end{equation}
For  small values of $n$ the biggest root of Eq. (\ref{equation1})
could be found analytically :\\
For $n = 1$, we have $\lambda = 1$
and topological entropy is equal to zero
and chaotic behavior is absent, that is clear from the physical point of view.\\
For $n = 2$,  we obtain
\begin{equation}
\lambda = \frac{1}{3}\left(\frac{27}{2} - \frac{3\sqrt{69}}{2}\right)
^{1/3} +\frac{\left(\frac{1}{2}(9+\sqrt{69})\right)^{1/2}}{3^{2/3}}
\approx 1.32 .
\label{m2}
\end{equation}
For $n = 3$, one can get
\begin{eqnarray}
&&\lambda = \frac{1}{3} +
\frac{1}{3}\left(\frac{29}{2}-\frac{3\sqrt{93}}{2}\right)^{1/3}\nonumber \\
&&+\frac{1}{3}\left(\frac{29}{2}+\frac{3\sqrt{93}}{2}\right)^{1/2}
\approx 1.47 .
\label{m3}
\end{eqnarray}
For higher values of $n$ one can find $\lambda$ numerically,
for example for $n = 4$, $\lambda \approx 1.53$.

For large values of $n$ one can find an asymptotical value for
the biggest root $\lambda$:
\begin{equation}
\lambda = \frac{1 + \sqrt{5}}{2} - \frac{1}{\sqrt{5}
\left(\frac{1 + \sqrt{5}}{2}\right)^{n}}.
\label{asymp}
\end{equation}

As it has already been mentioned above in the model with the scalar field
and matter or in the model with the complex scalar field and non-zero
classical charge \cite{complex}, the rules governing the structure are rather complicated.
Nevertheless in this case also it is possible to calculate the topological
entropy. Here we shall consider one particular example; however, the
algorithm, which shall be presented, could be used for different sets of
rules as well.

Thus, let us formulate the rules for our model:\\
1. It is impossible to have more than 19 letters $B$.\\
2. After a series with 19 letters $B$ (and letter $A$) one can have
the next series only with 1 letter $B$.\\
3. After a series with 18 letters $B$ one can have the next series
with 1 or 2 letters $B$.\\
4. After a series with 17 letter $B$ one can have the next series with
1, 2 or 3 letters $B$.\\
\ldots \\
5. After a series with 1 letter $B$  one can have the series with
$n$ letters $B$, where $0 \leq n \leq 19$.\\

Let us notice that the system of rules has a remarkable symmetry in
respect with number $n_{C} = 10$, which makes further calculations more
simple. These symmetrical rules give us a good approximation for
the description of the real physical situation. The value $n_{C}$ is apparently
a function of the parameters of the model under investigation.
More detailed numerical investigation implies
a more complicated system of rules, however, it also could be formalized
in the system of recurrent relations. Below we shall see that already
the symmetric system of rules gives rather a cumbersome equation
for the topological entropy.

Let us now introduce the following notation:\\
$Q(k)$ is the number of words which begin with the letter $A$
and end with the letter $A$,\\
$Q_{1}(k)$  is the number of words beginning with the letter $A$ and ending
with a series with 1 letter $B$,\\
$Q_{2}(k)$ is the number of words beginning with the letter $A$ and ending
with a series with 2 letters $B$,\\
\ldots,\\
$Q_{19}(k)$ is the number of words beginning with the letter $A$ and
ending with a series with 19 letters $B$.\\

The system of the recurrence relations for these quantities looks as follows:
\begin{eqnarray}
&&Q(k+1) = Q_{1}(k) + Q_{2}(k) + \cdots + Q_{19}(k),\nonumber \\
&&Q_{1}(k) = Q(k-1),\nonumber \\
&&Q_{19}(k) = Q_{1}(k-20),\nonumber \\
&&Q_{d}(k) = Q_{d-1}(k-1) - Q_{21-d}(k-d-1),\ 2 \leq d \leq 10
\nonumber \\
&&Q_{d}(k) = Q_{d+1}(k+1) + Q_{20-d}(k-d-1),\ 11 \leq d \leq 18
\label{req}
\end{eqnarray}
Resolving this system in respect with $Q(k)$, one get the following
recurrent relation:
\begin{eqnarray}
&&Q(k+1) = Q(k-1) + Q(k-2) + Q(k-3) + Q(k-4)\nonumber \\
&&+ Q(k-5) + Q(k-6) + Q(k-7) + Q(k-8) + Q(k-9)\nonumber \\
&&+ Q(k-10) + Q(k-13) + 2Q(k-14) + 3Q(k-15) + 4Q(k-16)\nonumber \\
&&+ 5Q(k-17) + 6Q(k-18) + 7Q(k-19) + 8Q(k-20) + 9Q(k-21)\nonumber \\
&&- 9Q(k-24) - 16Q(k-25) - 21Q(k-26) -24Q(k-27)\nonumber \\
&&- 25Q(k-28) - 24Q(k-29) - 21Q(k-30) - 16Q(k-31)\nonumber \\
&&- 9Q(k-32) - 8Q(k-36) - 21Q(k-37) - 36Q(k-38)\nonumber \\
&&- 50Q(k-39) - 60Q(k-40) - 63Q(k-41) - 56Q(k-42)\nonumber \\
&&- 36Q(k-43) + 36Q(k-47) + 84Q(k-48) + 126Q(k-49)\nonumber \\
&&+ 150Q(k-50) + 150Q(k-51) + 126Q(k-52) + 84Q(k-53)\nonumber \\
&&+ 36Q(k-54) + 28Q(k-59) + 84Q(k-60) + 150Q(k-61)\nonumber \\
&&+ 200Q(k-62) + 210Q(k-63) + 168Q(k-64) + 84Q(k-65)\nonumber \\
&&- 84Q(k-70) - 224Q(k-71) - 350Q(k-72) - 400Q(k-73)\nonumber \\
&&- 350Q(k-74) - 224Q(k-75) - 84Q(k-76) - 56Q(k-82)\nonumber \\
&&- 175Q(k-83) - 300Q(k-84) - 350Q(k-85) - 280Q(k-86)\nonumber \\
&&- 126Q(k-87) + 126Q(k-93) + 350Q(k-94) + 525Q(k-95)\nonumber \\
&&+ 525Q(k-96) + 350Q(k-97) + 126Q(k-98) + 70Q(k-105)\nonumber \\
&&+ 210Q(k-106) + 315Q(k-107) + 280Q(k-108) + 126Q(k-109)\nonumber \\
&&- 126Q(k-116) - 336Q(k-117) - 441Q(k-118) - 336Q(k-119)\nonumber \\
&&- 126Q(k-120) - 56Q(k-128) - 147Q(k-129) - 168Q(k-130)\nonumber \\
&&- 84Q(k-131) + 84Q(k-139) + 196Q(k-140) + 196Q(k-141)\nonumber \\
&&+ 84Q(k-142) + 28Q(k-151) + 56Q(k-152) + 36Q(k-153)\nonumber \\
&&- 36Q(k-162) - 64Q(k-163) - 36Q(k-164) - 8Q(k-174)\nonumber \\
&&- 9Q(k-175) + 9Q(k-185) + 9Q(k-186) + Q(k-197)\nonumber \\
&&- Q(k-208),
\label{req1}
\end{eqnarray}
which in turn gives the following equation for the topological entropy:
\begin{eqnarray}
&&x^{209} - x^{207} - x^{206} - x^{205} - x^{204} - x^{203} - x^{202}
- x^{201} - x^{200}\nonumber \\
&&- x^{199} - x^{198} - x^{195} - 2x^{194} - 3x^{193} - 4x^{192}
- 5x^{191} - 6x^{190}\nonumber \\
&&- 7x^{189} - 8x^{188} - 9x^{187} + 9x^{184} + 16x^{183}
+ 21x^{182} + 24x^{181}\nonumber \\
&&+ 25x^{180} + 24x^{179} + 21x^{178} + 16x^{177} + 9x^{176} + 8x^{172}
+ 21x^{171} \nonumber \\
&&+ 36x^{170} + 50x^{169} + 60x^{168} + 63x^{167} + 56x^{166} +
36x^{165} - 36x^{161}\nonumber \\
&&- 84x^{160} - 126x^{159} - 150x^{158} - 150x^{157} - 126x^{156}
- 84x^{155}\nonumber \\
&&- 36x^{154} - 28x^{149} - 84x^{148} - 150x^{147} - 200x^{146}
- 210x^{145} \nonumber \\
&&- 168x^{144} - 84x^{143} + 84x^{142} + 224x^{137} + 350x^{136}
+ 400x^{135}\nonumber \\
&&+ 350x^{134} + 224x^{133} + 84x^{132} + 56x^{126} + 175x^{125}
+ 300x^{124}\nonumber \\
&&+ 350x^{123} + 280x^{122} + 126x^{121} - 126x^{115} - 350x^{114}
- 525x^{113}\nonumber \\
&&- 525x^{112} - 350x^{111} - 126x^{110} - 70x^{103} - 210x^{102}
- 315x^{101}\nonumber \\
&&- 280x^{100} - 126x^{99} + 126x^{92} + 336x^{91} + 441x^{90}
+ 336x^{89}\nonumber \\
&&+ 126x^{88} + 56x^{80} + 147x^{79} + 168x^{78} + 84x^{77}
- 84x^{69}\nonumber \\
&&- 196x^{68} - 196x^{67} - 84x^{66} - 28x^{57} - 56x^{56}
- 36x^{55}\nonumber \\
&&+ 36x^{46} + 64x^{45} + 36x^{44} + 8x^{34} + 9x^{33}
- 9x^{23}\nonumber \\
&&- 9x^{22} - x^{11} + 1 = 0.
\label{equation3}
\end{eqnarray}
Resolving numerically Eq. (\ref{equation3}) one can find the biggest root
which is equal to $\lambda \approx 1.617 71$.

Correspondingly, the topological entropy is given by the logarithm of
the biggest root.

The scheme, described above, can be applied to many different physical
models, obeying the different sets of the prohibition rules governing
the structure of periodical trajectories.

\section{Quasi-isotropic expansion in cosmology}

The 
quasi-isotropic solution of the Einstein equations near a cosmological 
singularity was found by Lifshitz and Khalatnikov \cite{quasi} for the 
Universe filled by radiation  with the equation of state $p = 
\frac{\varepsilon}{3}$ in the early 60th. 
In the paper \cite{KKS}, there was presented the generalization of the 
quasi-isotropic solution of the Einstein equations near a cosmological 
singularity to the case of an arbitrary one-fluid 
cosmological model.
Then this solution was further generalized to the 
case of the Universe filled by two ideal barotropic fluids \cite{KKMS,KKS1}.

To explain the physical sense of the quasi-isotropic solution, let us 
remind that it represents the most generic spatially inhomogeneous 
generalization of the Friedmann spacetime in which the spacetime is locally
Friedmann-like near the cosmological singularity $t=0$ (in particular, its
Weyl tensor is much less than its Riemann tensor). On the other hand, 
generically it is very inhomogeneous globally and may have a very 
complicated spatial topology. As was shown in \cite{quasi,KKS} (see 
also \cite{Deruelle}), such a solution contains 3 arbitrary functions 
of space coordinates. From the Friedmann-Robertson-Walker (FRW) point of view, 
these 3 degrees of 
freedom represent the growing (non-decreasing in terms of metric 
perturbations) mode of adiabatic perturbations and the non-decreasing 
mode of gravitational waves (with two polarizations) in 
the case when deviations of a space-time metric from the FRW one are 
not small. So, the quasi-isotropic solution is not a generic solution 
of the Einstein equations with a barotropic fluid. Therefore, one 
should not expect this solution to arise in the course of generic 
gravitational collapse (in particular, inside a black hole event 
horizon). The generic solution near a space-like curvature singularity 
(for $p<\varepsilon$) has a completely different structure consisting 
of the infinite sequence of anisotropic vacuum Kasner-like eras with 
space-dependent Kasner exponents (see section 2 of the present review).

For this reason, the quasi-isotropic solution had not attracted 
much interest for about twenty years. Its new life began after the 
development of successful inflationary models (i.e., with "graceful 
exit" from inflation) and the theory of generation of perturbations
during inflation, because it had immediately become clear that 
generically (without fine tuning of initial conditions) the scalar metric
perturbations after the end of inflation remained small in a finite 
region of space which was much less than the whole causally connected 
space volume produced by inflation. It appears that the 
quasi-isotropic solution can be used for a global description of a 
part of space-time after inflation, which belongs to "one 
post-inflationary universe". The latter is defined as a connected part 
of space-time, where the hyper-surface $t=t_f({\bf r})$ describing the 
moment when inflation ends is space-like and, therefore, can be made 
the surface of constant (zero) synchronous time by a coordinate 
transformation. This directly follows from the derivation of 
perturbations, generated during inflation given in \cite{Star82} (see 
Eq. (17) of that paper) which is valid in case of large perturbations, 
too. Thus, when used in this context, the quasi-isotropic solution 
represents an {\em intermediate} asymptotic regime during expansion of 
the Universe after inflation. The synchronous time $t$, appearing in it,
is the proper time since the {\em end} of inflation, and the region 
of validity of the solution is from $t=0$ up to a moment in future 
when spatial gradients become important. For sufficiently large 
scales, the latter moment may be rather late, even of the order or
larger than the present age of the Universe. Note also  the analogue 
of the quasi-isotropic solution {\em before} the end of inflation is 
given by the generic quasi-de Sitter solution found in \cite{Star83}.
Both solutions can be smoothly matched across the hypersurface of 
the end of inflation.

A slightly different versions of the quasi-isotropic expansion was developed 
during the last decades which are known under the names of long-wave expansion or gradient expansion 
\cite{long-wave}. 

Originally the quasi-isotropic expansion was developed as a technique of generation of some kind of perturbative expansion 
in the vicinity of the cosmological singularity, where the cosmic time parameter served as a perturbative one. However,
the more general treatment of the quasi-isotropic expansion is possible if one notices that the next order of the 
quasi-isotropic expansion contain the higher orders of the spatial derivatives of the metric coefficients. 
Thus, it is possible to construct a natural generalization of the quasi-isotropic solution of the Einstein equations 
which would be valid not only in the vicinity of the cosmological singularity, but in the full time range.
In this case the simple algebraic equations, which one resolves to find the higher orders of the quasi-isotropic 
approximation in the vicinity of the singularity are substituted by differentially equations, where the time dependence 
of the metric can be rather complicated in contrast to the power-law behavior of the coefficients of the original 
quasi-isotropic expansion.

Thus, we now show how does it work for a universe, filled with one barotropic fluid \cite{KKS1}. 
We consider the universe with the fluid with the equation of state $p = w\varepsilon$. The spatial metric 
now is 
\begin{equation}
\gamma_{\alpha\beta} = a_{\alpha\beta}t^{\kappa} + c_{\alpha\beta},
\label{metricA}
\end{equation}
where
\begin{equation}
\kappa = \frac{4}{3(1+w)}.
\label{kappa}
\end{equation}
Inverse metric is
\begin{equation}
\gamma^{\alpha\beta} = a^{\alpha\beta}t^{-\kappa} - c^{\alpha\beta}t^{-2\kappa}.
\label{inverseA}
\end{equation}
Then we have the following formulae for the extrinsic curvature:
\begin{equation}
K_{\alpha\beta} = a_{\alpha\beta} \kappa t^{\kappa -1} + \dot{c}_{\alpha\beta},
\label{extrinsicA}
\end{equation}
\begin{equation}
K_{\alpha}^{\beta} = \delta_{\alpha}^{\beta} \kappa + \dot{c}_{\alpha}^{\beta}t^{-\kappa}-c_{\alpha}^{\beta}\kappa t^{-\kappa-1},
\label{extrinsicA1} 
\end{equation}
\begin{equation}
K = \frac{3\kappa}{t} + \dot{c}t^{-\kappa}-c\kappa t^{-\kappa-1},
\label{extrinsicA2}
\end{equation}
\begin{equation}
\frac{\partial K}{\partial t} = -\frac{3\kappa}{t^2} + \ddot{c}t^{-\kappa} - 2\dot{c}\kappa t^{-\kappa-1} + c\kappa(\kappa+1) t^{-\kappa-2},
\label{extrinsicA3}
\end{equation}
\begin{equation}
K_{\alpha}^{\beta}K_{\beta}^{\alpha}  = \frac{3\kappa^2}{t^2} + 2\dot{c}\kappa t^{-\kappa-1} - 2c\kappa^2t^{-\kappa-2}.
\label{extrinsicA4}
\end{equation}
Using formulae (\ref{extrinsicA3}), (\ref{extrinsicA4}), we have
\begin{equation}
R_{0}^{0} = \frac{3\kappa(2-\kappa)}{4t^2} - \frac{\ddot{c}t^{-\kappa}}{2} + \frac{\dot{c} \kappa t^{-\kappa-1}}{2} - \frac{c\kappa t^{-\kappa-2}}{2}.
\label{R00A}
\end{equation}
Using now the Einstein equation  in the lowest order of the approximation we obtain for the energy density of the fluid under consideration 
\begin{equation}
\varepsilon^{(0)} = \frac{4}{3(1+w)^2t^2}.
\label{energy0A} 
\end{equation}
Using the 0 component of the energy-momentum conservation law  we can find the relation between 
the first correction to the energy density $\varepsilon^{(1)}$ and the trace of the first correction to the metric $c$:
\begin{equation}
\varepsilon^{(1)} = -\frac{c\kappa t^{-\kappa-2}}{2}.
\label{energyA1}
\end{equation}
Now, using the standard expression for the scalar curvature $R$  and the  00 component of the Einstein equation in the form $R_0^0 - \frac12 R = T_0^0$ for in the first quasi-isotropic order the following equation:
\begin{equation}
\frac{P}{2} + \frac14 K^{(0)}K^{(1)} - \frac18 (K_{\alpha}^{\beta}K_{\beta}^{\alpha})^{(1)} = \varepsilon^{(0)}.
\label{dif-eqA} 
\end{equation}
Combining (\ref{dif-eqA}) and (\ref{energyA1}) we obtain the following differential equation for $c$:
\begin{equation}
\dot{c} = \frac{c (\kappa -1)}{t} - \frac{\bar{P} t}{\kappa}.
\label{dif-eqA1}
\end{equation}
Integrating (\ref{dif-eqA1}) we obtain
\begin{equation}
c = \frac{\bar{P}t^2}{\kappa(\kappa-3)} = -\frac{9(1+w)^2\bar{O}t^2}{4(5+9w)}.
\label{cA}
\end{equation}

Now to find the traceless part of the first correction to the metric $\tilde{c}_{\alpha\beta}$ we shall use the traceless 
part of the spatial-spatial component of the Einstein equations, which in the case of one fluid and in the first order approximation has a particularly simple form:
\begin{equation}
\tilde{R}_{\alpha}^{\beta} = -\tilde{P}_{\alpha}^{\beta} - \frac12\frac{\partial \tilde{K}_{\alpha}^{\beta}}{\partial t} -\frac{3\kappa}{4t}\tilde{K}_{\alpha}^{\beta} = 0.
\label{tracelessA}
\end{equation}
(Notice that the traceless part of the extrinsic curvature $\tilde{K}_{\alpha}^{\beta}$ does not have zero-order terms).
The equation (\ref{tracelessA}) can be rewritten as 
\begin{equation}
\frac{\partial \tilde{K}_{\alpha}^{\beta}}{\partial t} + \frac{3\kappa}{2t}\tilde{K}_{\alpha}^{\beta} = 
-2\tilde{\bar{P}}_{\alpha}^{\beta} t^{-\kappa}.
\label{tracelessA1}
\end{equation}
Integrating (\ref{tracelessA1}) one obtains
\begin{equation}
\tilde{K}_{\alpha}^{\beta} = -\frac{4}{\kappa+2}\tilde{\bar{P}}_{\alpha}^{\beta} t^{-\kappa+1}
\label{tracelessA2} 
\end{equation}
Using the relation
\begin{equation}
\tilde{c}_{\alpha}^{\beta} = t^{\kappa}\int \tilde{K}_{\alpha}^{\beta} dt,
\label{relationA} 
\end{equation}
we come to 
\begin{equation}
\tilde{c}_{\alpha}^{\beta} = \frac{4}{\kappa^2-4}\tilde{\bar{P}}_{\alpha}^{\beta} t^2 = 
-\frac{9(1+w)^2}{(3w+1)(3w+5)}\tilde{\bar{P}}_{\alpha}^{\beta} t^2.
\label{tracelessA3}
\end{equation}

One can see that the results (\ref{cA}) and (\ref{tracelessA3}) valid in the full range of time coincide with 
those valid in the vicinity of the initial cosmological singularity ($t = 0$) \cite{KKS} obtained by the algebraic method 
\cite{quasi}. 
The general expression for the first correction to the metric for one-fluid case is given in the formula (37) in \cite{KKS}.
The metric $b_{\alpha\beta}$ in \cite{KKS} corresponds to $c_{\alpha\beta}$ in the present paper, while for the equation of 
state parameter the symbol $k$ is used instead of $w$.  
The formula (37) contains a misprint: in front of the second term in the brackets in the right-hand side of this equation 
should stay the factor $1/4$. At first glance the first correction to the metric (37) contains a pole at $3k+1 = 0$, however
calculating the trace of this metric, one sees that this pole is cancelled and is present  only in its  anisotropic part. 

Thus, for the case of the universe filled with the  string gas $w = -\frac13$ the quasi-isotropic expansion does not 
work because the expression for $\tilde{c}_{\alpha\beta}$ becomes singular. 

In the conclusion let us consider a special case when the metric $a_{\alpha\beta}$ has a conformally flat form:
\begin{equation}
a_{\alpha\beta} = e^{\rho(x)}\delta_{\alpha\beta}.
\label{flat}
\end{equation}
In this case the spatial Ricci tensor is 
\begin{equation}
\bar{P}_{\alpha\beta} = \frac14(\rho_{,\alpha}\rho_{\beta} - 2 \rho_{,\alpha\beta}) - \frac14\delta_{\alpha\beta}(\rho_{,}^{\mu}
\rho_{,\mu} + 2 \rho_{,\mu}^{\mu})
\label{flat1}
\end{equation}
or 
\begin{equation}
\bar{P}_{\alpha}^{\beta} = \frac14(\rho_{,\alpha}\rho_{,}^{\beta} - 2 \rho_{,\alpha}^{\beta}) -\frac14\delta_{\alpha}^{\beta}
(\rho_{,}^{\mu}\rho_{,}^{\mu} + 2\rho_{,\mu}^{\mu}).
\label{flat2}
\end{equation}
Correspondingly
\begin{equation}
\bar{P} = -2\rho_{,\mu}^{\mu} -\frac12\rho_{,\mu}^{\mu}
\label{flat3}
\end{equation}
and the traceless part of the Ricci tensor is 
\begin{equation}
\tilde{\bar{P}}_{\alpha}^{\beta} = \frac14 (\rho_{,\alpha}\rho_{,}^{\beta} - 2\rho_{,\alpha}^{\beta}) 
+ \frac{1}{12} \delta_{\alpha}^{\beta}(2\rho_{,\mu}^{\mu} - \rho_{,\mu}\rho_{,}^{\mu}).
\label{flat4}
\end{equation}
If 
\begin{equation}
\rho = A_{\mu\nu}x^{\mu}x^{\nu}
\label{Gauss}
\end{equation}
then 
\begin{equation}
\bar{P} = -4A_{\mu}^{\mu} - 2A_{\mu\nu}A_{\alpha}^{\mu}x^{\nu}x^{\alpha}
\label{Gauss1}
\end{equation}
and 
\begin{equation}
\tilde{\bar{P}}_{\alpha}^{\beta} = \frac13 x^{\gamma}x^{\nu}(3A_{\alpha\gamma}A_{\nu}^{\beta} - \delta_{\alpha}^{\beta} A_{\mu\gamma}
A_{\nu}^{\mu}).
\label{Gauss2}
\end{equation}

Thus, it is easy to see that if the metric in the lowest order of the quasi-isotropic expansion has the Gaussian form 
determined by Eqs. (\ref{flat}) and (\ref{Gauss}) already its first correction $c_{\alpha\beta}$ determined by the 
curvature tensors (\ref{Gauss1}) and (\ref{Gauss2}) has non-Gaussian form due to the presence of the quadratic 
in $x^{\mu}$ terms in front of the Gaussian exponential.

\section{Quantum field theory and quantum  gravity at a Lifshitz point}

It is generally recognized that the complete theory of elementary particles and fundamental interactions should include quantum gravity. However, the quantum gravity theory is non-renormalizable (see e.g. \cite{Weinberg}). The main obstacle again perturbative renormalizability of the general relativity in $3+1$ dimensions is the fact that the gravitational coupling constant (the Newton constant) is dimensionful with a negative dimension $[G_N]=-2$ in mass units. The graviton propagator as all the propagators in quantum field theory scales with the four-momentum $k_{\mu}=(\omega,\vec{k})$ as 
\begin{equation}
\frac{1}{k^2},
\label{Hor}
\end{equation}
where $k=\sqrt{\omega^2-\vec{k}^2}$. When we calculate Feynman diagrams with an increasing number of loops, it is necessary to introduce again and again the counterterms with increasing degree in curvature. 

An improved ultraviolet behavior of the theory can be obtained if some higher-order in curvatures terms are added to the Lagrangian. The terms quadratic in curvature not only yield new interactions (with a dimensionless coupling), but also modify the propagator. 
Omitting the tensor structure of this propagator, one can write it as 
\begin{eqnarray}
&&\frac{1}{k^2}+\frac{1}{k^2}G_Nk^4\frac{1}{k^2}+\frac{1}{k^2}G_Nk^4\frac{1}{k^2}G_Nk^4\frac{1}{k^2}+\cdots\nonumber \\
&&=\frac{1}{k^2-G_Nk^4}.
\label{Hor1}
\end{eqnarray}
It is easy to see that at high energies the propagator is dominated by $1/k^4$ term. This cures the problem of ultraviolet divergences. However, a new problem arises: The resummed propagator (\ref{Hor1}) has two poles:
\begin{equation}
\frac{1}{k^2-G_Nk^4}= \frac{1}{k^2}-\frac{1}{k^2-1/G_N}.
\label{Hor2}
\end{equation}
One of these poles describes massless gravitons, while the other corresponds to ghost excitations and implies violations of unitarity.  

Some years ago P Horava has introduced a new class of gravity models \cite{Hor-Lif},
which he has called ``Quantum gravity at a Lifshitz point''. The creation of this approach was inspired by the papers E M Lifshitz \cite{Lif-phase}, devoted to the theory of second-order phase transitions and of critical phenomena and published in 1941. The Horava-Lifshitz gravity models exhibit scaling properties which are anisotropic between space and time. The degree of anisotropy between space and time is measured by the dynamical critical exponent $z$ such that 
\begin{equation}
\vec{x} \rightarrow b\vec{x},\ \ t \rightarrow b^z t.
\label{z-crit}
\end{equation}
Such an anisotropic scaling is common in condensed matter systems. The prototype of the class of the condensed matter models of this family is the theory of a Lifshitz scalar in $D+1$ dimensions , whose action is  
\begin{equation}
S = \int dt d^Dx\{(\dot{\Phi})^2-(\triangle \Phi)^2\},
\label{Lif-ac}
\end{equation}
where $\triangle$ is the spatial Laplacian. Here the critical exponent $z=2$. We can add to the action a term 
\begin{equation}
-c^2\int dt d^Dx\partial_i\Phi\partial_i\Phi,
\label{Lif-ac1}
\end{equation}
where we have introduced explicitly the velocity of light $c$. 
Under the influence of this deformation, the theory flows in the infrared region to the theory with $z=1$, with the Lorentz invariance emerging at long distances. 

In the approach to the quantum gravity, suggested in \cite{Hor-Lif}, there were considered actions such that scaling at short distances exhibited a strong anisotropy between space and time, with $z > 1$. This improves the short-distance behavior of the theory. Indeed, the propagator for such gravitons is proportional to 
\begin{equation}
\frac{1}{\omega^2-c^2\vec{k}^2-G(\vec{k}^2)^z}.
\label{prop-Lif}
\end{equation}

At high energies the propagator is dominated by the anisotropic term $1/(\omega^2-G(\vec{k}^2)^z$. For a suitably chosen $z$, this modification improves the short-distance behavior and the theory becomes power-counting renormalizable. The $c^2\vec{k}^2$ term becomes important at low energies, where the theory naturally flows to $z=1$. 

Unlike in relativistic higher-derivative theories mentioned above, higher-order time derivatives are not generated, and the problem with ghost excitations and non-unitarity is resolved. In the case of the $3+1$ gravity, the renormalizability is achieved by the choice of $z=3$. 

Now we consider the form of the Lagrangian of the gravitational theory respecting the symmetry (\ref{z-crit}). We shall introduce as usual the $D+1$ split of the manifold under consideration with the spatial metric tensor $g_{ij}$ and the lapse $N$ and shift $N_i$ functions \cite{ADM}. 
We would like to construct the kinetic term of the Lagrangian which is quadratic in time derivatives $\dot{g}_{ij}$ and invariant under the foliation-preserving diffeomorphisms, i.e. diffeomorphisms, which respect $D+1$ foliation. Such kinetic term  depends on the second fundamental form (extrinsic curvature):
\begin{equation}
K_{ij}=\frac{1}{2N}(\dot{g}_{ij}-\nabla_iN_j-\nabla_jN_i),
\label{ext-curv}
\end{equation}
where $\nabla_i$ is the covariant derivative, involving the spatial metric. 
Then, the kinetic part of the action is 
\begin{equation}
S_K=\frac{2}{\kappa^2}\int dtd^Dx\sqrt{g}N(K_{ij}K^{ij}-\lambda K^2).
\label{kin-Hor}
\end{equation}
In general relativity, the requirement of the of invariance under all spacetime diffeomorphisms   implies $\lambda=1$. 

The potential term of the action should include only spatial metric and its spatial derivatives and it has the form 
\begin{equation}
S_V=\int dtd^Dx\sqrt{g}NV[g_{ij}].
\label{pot-Hor}
\end{equation}

Considering the high-energy regime of the theory we should concentrate on the terms, which have the dimensionality as the kinetic term. In the case of $D=3$ and $z=3$ there are many terms of this kind. Some of these terms are quadratic in curvature 
\begin{equation}
\nabla_k R_{ij}\nabla^kR^{ij},\ \nabla_kR_{ij}\nabla^iR^{jk},\ R\triangle R, \ R^{ij}\triangle R_{ij};
\label{pot-Hor1}
\end{equation}
they modify the propagator. Other terms, such as 
\begin{equation}
R^3,\ \ R_j^iR_k^jR_i^k,\ \ RR_{ij}R^{ij},
\label{pot-Hor2}
\end{equation}
are cubic in curvature and represent pure interacting terms. The list of independent operators is very large, implying proliferation of coupling constants. In order to reduce the number of independent coupling constants, it is necessary to impose an additional symmetry on the theory. The way in which this restriction is implemented is very similar to that, which is used in the theory of critical phenomena. 

We will require the potential term to be of special form
\begin{equation}
S_V=\frac{\kappa^2}{8}\int dtd^Dx\sqrt{g}NE^{ij}G_{ijkl}E^{kl},
\label{pot-Hor3}
\end{equation}
where the tensor $E^{ij}$ itself follows from a variation 
\begin{equation}
\sqrt{g}E^{ij}=\frac{\delta W[g_{kl}]}{\delta g_{ij}},
\label{pot-Hor4}
\end{equation}
of some action $W$, while the tensor $G_{ijkl}$ is the inverse of the generalized DeWitt supermetric 
\begin{equation}
G^{ijkl}= \frac12(g^{ik}g^{jl}+g^{il}g^{jk})-\lambda g^{ij}g^{kl}.
\label{DeWitt}
\end{equation}

They say that the theories whose potential is of the form (\ref{pot-Hor3}) for some $W$ satisfy the ``detailed balance condition''. The systems which satisfy the detailed balance condition have simpler quantum behavior than general systems. Their renormalization can be reduced to the simpler renormalization of the associated theory described by $W$. 

We are interested in constructing a theory which satisfies detailed balance, and exhibits the short-distance scaling with $z=3$ leading to power-counting renomalizability in $3+1$ dimensions. Therefore, $E^{ij}$ must be of third order in spatial derivatives.  It turns out that there is a unique tensor with necessary properties: it is the Cotton tensor
\begin{equation}
C^{ij}=\varepsilon^{ikl}\nabla_k\left(R_l^j-\frac14R\delta_l^j\right), 
\label{Cotton}
\end{equation}
which is the variation of the action 
\begin{equation}
W=\int d^3x\varepsilon^{ijk}\left(\Gamma_{il}^m\partial_j\Gamma_{km}^l+\frac23\Gamma_{il}^n\Gamma_{jm}^l\Gamma_{kn}^m\right).
\label{Cotton1}
\end{equation}

Now we can write down the full action for $z=3$ gravity theory in $3+1$ dimensions:
\begin{eqnarray}
&&S=\int dtd^3x\sqrt{g}N\left\{\frac{2}{\kappa^2}(K_{ij}K^{ij}-\lambda K^2)-\frac{\kappa^2}{2w^4}C_{ij}C^{ij}\right\}\nonumber \\
&&=\int dtd^3x\sqrt{g}N\left\{\frac{2}{\kappa^2}(K_{ij}K^{ij}-\lambda K^2)-\frac{\kappa^2}{2w^4}\right.\nonumber \\
&&\left.\times\left(\nabla_iR_{jk}\nabla^iR^{jk}-\nabla_iR_{jk}\nabla^jR^{ik}-\frac18\nabla_iR\nabla^iR\right)\right\}.
\label{Cotton2}
\end{eqnarray}
This action depends on 3 constants ($\kappa,w$ and $\lambda$) and some relevant terms, providing the correct infrared limit of the theory can be added to it. It was shown that at special values of the parameter $\lambda=1/3$ or $\lambda = 1$, the metric field has 2 degrees of freedom just like the graviton in the general relativity. 

The relevant terms can respect the detailed balance condition. One can add to the action $W$ the term 
\begin{equation}
\mu\int d^3x \sqrt{g}(R-2\Lambda_W). 
\label{Cotton3}
\end{equation}
In this case the action shall have the form
\begin{eqnarray}
&&S=\int dtd^3x\sqrt{g}N\left\{\frac{2}{\kappa^2}(K_{ij}K^{ij}-\lambda K^2)-\frac{\kappa^2}{2w^4}C_{ij}C^{ij}\right.\nonumber \\
&&+\frac{\kappa^2\mu}{2w^2}\varepsilon^{ijk}R_{il}\nabla_jR_k^l-\frac{\kappa^2\mu^2}{8}R_{ij}R^{ij}\nonumber \\
&&\left.+\frac{\kappa^2\mu^2}{8(1-3\lambda)}\left(\frac{1-4\lambda}{4}R^2+\Lambda_WR-2\Lambda_W^2\right)\right\}.
\label{Cotton4}
\end{eqnarray}
At long distances, the potential is dominated by the last two terms in (\ref{Cotton4}): the spatial curvature scalar and the constant term. As a result the theory flows in the infrared region to $z=1$. 

Summing up, one can say that being inspired by  old Lifshitz papers \cite{Lif-phase} 
P Horava has invented a new quantum gravity theory, where the Lorentz invariance arises only at relatively long distances, while in the ultraviolet limit it is  absent. This theory is renormalizable and does not suffer from the presence of ghosts. 

The idea to break the Lorentz invariance at very small distances to provide the ultraviolet 
renormalizability of the theories with higher-derivative terms was used also outside the gravity context. In paper \cite{Lif-burst} the renormalizable field theories, including the scalar fields and Yang-Mills fields with higher-derivative terms in action, were considered. The presence of  higher-derivative terms changes the respective dispersion relations and  
the effective speed of life can grow to infinity in the ultraviolet limit of the theory. This has permitted to the authors of \cite{Lif-burst} to explain observable time delays in the gamma-ray bursts. 

Rather a large amount of work was connected with applications of the Horava-Lifshitz gravity to cosmology (for a review see \cite{Hor-cosm}). There are a number of interesting cosmological implications of the Horava-Lifshitz gravity. Let us dwell on some of them. 

First, let us notice that the action of Horava-Lifshitz gravity is invariant with respect to the space-time dependent spatial diffeomorphisms and with respect to only time-dependent time reparametrizations. That means that instead of 4 local first-class constraints, which we have 
in the general relativity (corresponding to variation with respect to lapse and shift functions), we have three local constraints, corresponding to three components of the shift function and a global constraint corresponding to the lapse function. The absence of the local constraint means that the analog of $00$ component of the Einstein equations is not valid \cite{Hor-cosm}. If we consider a Friedmann universe, that means that we cannot use the first Friedmann equation and we should use the second Friedmann equation, including the time derivative of the Hubble parameter $H = \dot{a}/a$. This equation has a first integral, which gives us the first Friedmann equation, but including   an additional constant. The appearance of this constant is equivalent to the appearance of the substance, which gravitationally behaves just as a dust-like matter. Thus, we have some kind of dark matter without dark matter. This effect is present already in the  flat Friedmann model.  

Now, following the paper \cite{Hor-cosm}, we shall write down the general action for the $z=3$ Horava-Lifshitz gravity, without thinking of the detailed balance condition and including the lower-derivative terms. This action has the following form:
\begin{eqnarray}
&&I = I_{kin}+I_{z=3}+I_{z=2}+I_{z=1}+I_{z=0}+I_m,\nonumber \\
&&I_{kin}=\frac{1}{16\pi G}\int Ndt\sqrt{g}d^3x(K^{ij}K_{ij}-\lambda K^2),\nonumber \\
&&I_{z=3}=\int Ndt\sqrt{g}d^3x(c_1\nabla_iR_{jk}\nabla^jR^{jk}+c_2\nabla_iR\nabla^iR+
c_3 R_i^jR_j^kR_k^i+c_4RR_i^jR_j^i+c_5R^3,\nonumber \\
&&I_{z=2}=\int Ndt\sqrt{g}d^3x(c_6R_i^jR_j^i+c_7R^2),\nonumber \\
&&I_{z=1}=c_8\int Ndt\sqrt{g}d^3x R,\nonumber  \\
&&I_{z=0}=c_9\int Ndt\sqrt{g}d^3x,
\label{hor-cosm-ac}
\end{eqnarray}
where $I_m$ is the matter action. 

For the Friedmann universe with an arbitrary curvature $k$, the second Friedmann equation for such a theory with the action (\ref{hor-cosm-ac}) is 
\begin{equation}
-\frac{3\lambda-1}{2}(2\dot{H}+3H^2)=8\pi G P -\frac{\alpha_3 k^3}{a^6}-\frac{\alpha_2 k^2}{a^4}+\frac{k}{a^2}-\Lambda,
\label{Hor-Fried}
\end{equation}
where 
\begin{equation}
\alpha_3=192\pi G(c_3+3c_4+9c_5),\ \ \alpha_2=32\pi G(c_6+3c_7),
\label{Hor-Fried1}
\end{equation}
where $P$ is the matter pressure. 
The first integral of this equation is 
\begin{equation}
\frac{3(3\lambda-1)}{2}H^2=8\pi G\left(\rho+\frac{C}{a^3}\right)-\frac{\alpha_3 k^3}{a^6}-
\frac{3\alpha_2k^2}{a^4}-\frac{3k}{a^2}+\Lambda,
\label{Hor-Fried2}
\end{equation}
where $\rho$ is the energy density of matter and $C$ is the effective dark matter integration constant, mentioned above. 

Now it is convenient to rewrite Eq. (\ref{Hor-Fried2}) as the energy conservation equation 
\begin{equation}
\frac{\dot{a}^2}{a^2}+\frac{2}{3\lambda-1}V(a)=0,
\label{Hor-Fried3}
\end{equation}
where 
\begin{equation}
V(a) =\frac{\alpha_3 k^3}{6a^4}+\frac{\alpha_2 k^2}{2a^2}+\frac{k}{2}-\frac{\Lambda}{6}a^2-\frac{4\pi G}{3}\left(\rho a^2+\frac{C}{a}\right).
\label{Hor-Fried4}
\end{equation}
The shape of the potential $V(a)$ completely determines the behavior of the system. 
When $\lambda > \frac13$ the universe can have only such values of $a$ that the potential $V(a) \leq 0$. In the points where $V(a) = 0$ the universe has turning points, i.e. the points of minimal contraction (bounce) or the points of maximal expansion.  

Thus, choosing appropriately the coefficients at the higher order spatial curvature terms, one can have different  cosmological regimes. We shall give some interesting examples.\\
1. If there is only one value of the cosmological radius $a=a_0$, where the potential $V(a)$ is equal to zero and if at $a < a_0$ the potential $V(a) > 0$ and at $a > a_0$ the potential 
$V(a) < 0$, then the universe has a bounce solution. Namely, at the beginning of the evolution the universe is contracting and after the bounce it is expanding. \\
2. If  there  are two values $a_1$ and $a_2$ such that $V(a_1)=V(a_2)=0$, and $V(a) < 0$ at $a_1 < a < a_2$, and $V(a) > 0$ at $a>a_2$ or $a<a_1$, then the universe has both the point of minimal contraction and the point of maximal expansion and its evolution is periodic.\\
3. If $V(a_0) = 0$ and in this point the function $V(a)$ has a local maximum, one has a non-stable static universe.\\
4. If at the point $a_0$ the function $V(a)$ has a local minimum and is equal to zero, we have a stable static universe.\\

The Horava-Lifshitz cosmology has another interesting feature. In its framework it is possible to predict the scale-invariant cosmological perturbations without requiring the existence of an inflationary stage of its expansion \cite{Hor-cosm}. Let us remember that the dispersion relation  for the standard linear cosmological perturbations is 
\begin{equation}
\omega^2=c_s^2\frac{k_c^2}{a^2},
\label{disp}
\end{equation}
where $c_s$ is the sound speed and $k_c$ is the comoving wave number. 

If a mode under consideration satisfies $\omega^2 \gg H^2$, where $H=\frac{\dot{a}}{a}$ is the Hubble parameter then the evolution of this model does not feel the expansion of the universe and the mode simply oscillate.  When $\omega^2 \ll H^2$, the expansion of the universe is so rapid that the Hubble friction freezes the model, which remains almost constant. Generation of cosmological perturbations from quantum fluctuations is the 
oscillation followed by the freeze-out. Therefore, the condition for the generation of cosmological perturbations is 
\begin{equation}
\frac{d}{dt}\left(\frac{H^2}{\omega^2}\right) > 0.
\label{disp1}
\end{equation}
If the standard dispersion relation (\ref{disp}) is valid, Eq. (\ref{disp1}) implies $\ddot{a} > 0$
for expanding universe. Therefore, the generation of cosmological perturbations from quantum fluctuations requires accelerated expansion of the universe, i.e. inflation. 

In the Horava-Lifshitz gravity with the critical index $z$, the dispersion relation for the perturbations in the ultraviolet region is
\begin{equation}
\omega^2 = M^2\times\left(\frac{k_c^2}{M^2a^2}\right)^z,
\label{disp2}
\end{equation}
where $M$ is some energy scale. By 
substituting Eq. (\ref{disp2}) into the condition (\ref{disp1}) we obtain 
\begin{equation}
\frac{d^2 a^z}{dt^2} > 0 
\label{disp3}
\end{equation}
for an expanding universe. Obviously, for large values of $z$, for example for $z =3$,
the condition (\ref{disp3}) does not require an accelerating universe. Indeed, already power-law expansion $a \sim t^p$, where $p > 1/z$ does the job. Thus, in the Horava-Lifshitz cosmology, one can generate the spectrum of cosmological perturbations without inflation.

Concluding this section, we would like to add that practically all traditional aspects of cosmology were also studied in the context of the Horava-Lifshitz gravity. The theory of linear perturbations, ascending to the pioneering work by Lifshitz of 1946 \cite{Lif1946} was studied, for example, in papers \cite{Hor-pert}. 
The quasi-isotropic (gradient) expansion for the Horava-Lifshitz cosmology in the presence 
of the scalar field was considered in paper \cite{Hor-quasi}.
There was also some activity concerning the oscillatory approach to the singularity and the stochasticity phenomena in Horava-Lifshitz cosmology \cite{Hor-mix}.
It looks like  the question about the presence or absence of such stochasticity is not yet resolved and requires further studies.

\section{Conclusions}

In this review we have considered some new directions of development of theoretical physics, connected with the gravitation and cosmology, which in one or another way stem 
from some old works by E M Lifshitz. Namely, we have dwelled on such topics as 
oscillatory approach to the singularity and stochasticity in the anisotropic and isotropic cosmology, quasi-isotropic expansion, and the so-called Horava-Lifshitz gravity and cosmology. 

Sometimes, these developments appear to be rather unexpected and amusing. Indeed, the study of the oscillatory approach to the cosmological singularity and stochasticity phenomena, being extended to cosmological models, based on superstring theories, revealed connections between the cosmological billiard dynamics and the properties of the infinite-dimensional Kac-Moody algebras.  

The ideas of some anisotropy between space and time, useful for the  description of the second-order phase transitions and critical phenomena, contained in the paper by Lifshitz, written in 1941, inspired the creation of the Horava-Lifshitz gravity, which looks to be renormalizable at the quantum level and which being applied to the cosmology, shows a lot of interesting effects and properties. 

Now, coming to the end of our paper, we would like to mention briefly one more paper by Lifshitz, which is connected with quantum field theory, gravity and condensed matter physics 
simultaneously. We mean the theory of macroscopic Van-der-Waals forces between solids 
\cite{Lif-W}. It is well-known that in quantum field theory one always has  the vacuum energy which diverges. Considering standard scattering processes one can forget about this energy,
subtracting it by means of the operation of the normal ordering. In 1948 Casimir has understood, that if the electromagnetic field satisfies some special boundary conditions, then its vacuum energy is different from that of the Minkowski spacetime and the difference between the correspondent energy densities can be finite \cite{Casimir}. Namely, he considered two infinite conducting planes and has shown that this difference was negative and that it implied the presence of the attracting force. This effect, called the Casimir effect, was also observed experimentally. Later, the role of the Casimir energy in cosmology was studied by different authors (see e.g. \cite{Ford}). 

In his paper published in 1956 \cite{Lif-W} E M Lifshitz has studied another aspect of the Casimir effect. Namely, Casimir has considered an ideal conductor, providing absolute screening of the electromagnetic field. Thus, the tangential component of the electric field disappears on the conducting planes. Lifshitz, instead, has noticed that the ideal conductors do not exist and that the dielectric permittivity (which is infinite for the ideal conductor) really depends on the frequency of the electromagnetic field and for higher frequencies this characteristics tends to one and the material becomes transparent for electromagnetic fields. The formulae taking into account this effect and also the thermic fluctuations together with quantum ones, have created the basis for a new direction in physics, which sometimes is called the theory of macroscopic Van-der-Waals forces \cite{Lif-W1}. 
This approach to the Casimir effect is also very popular now and the corresponding forces which take their origin from quantum and thermic fluctuations are called Casimir-Lifshitz forces \cite{Cas-Lif}.

This work was partially supported by the RFBR grant 14-02-00894.

\end{document}